\documentclass[12pt,letterpaper]{article}
\usepackage{epsfig,graphicx}
\usepackage{setspace,multirow}
\usepackage{amsmath,amsthm}
\usepackage{amssymb,enumerate}
\usepackage{bbm}
\usepackage[noend]{algorithmic}
\usepackage{setspace,paralist,color}
\usepackage{bm,verbatim,subfigure}
\usepackage{subfigure}
\usepackage{lineno}
\usepackage{multicol}
\usepackage{rotating}
\usepackage{threeparttable}
\usepackage{CJK}
\usepackage[titletoc]{appendix}
\usepackage{titlesec}
\usepackage[round]{natbib}
\usepackage{hyperref}
\hypersetup{hypertex=true,
	colorlinks=true,
	linkcolor=blue,
	anchorcolor=blue,
	citecolor=blue}

\allowdisplaybreaks
\def\wt{\widetilde}
\renewcommand{\hat}{\widehat}
\def\be{\begin{eqnarray}}
\def\ee{\end{eqnarray}}
\def\argmin{\mathrm{argmin}}
\def\argmax{\mathrm{argmax}}

\def\cov{\mathrm{Cov}}

\newcommand{\E}{{\mathbb{E}}}
\def\n{\nonumber}

\def\sumi{\sum_{i=1}^n}
\def\summ{\sum_{m=1}^M}

\def\var{\mathrm{Var}}

\def\wt{\widetilde}
\def\tt{^{\top}}

\def\w{{\bf w}}
\def\X{{\bf X}}
\def\Xn{{\bf X}_{\mathrm{new}}}
\def\x{{\bf x}}
\def\I{{\bf I}}

\def\Y{{\bf Y}}

\def\calW{{\mathcal{W}}}
\def\calL{{\mathcal{L}}}
\def\calU{{\mathcal{U}}}
\def\C{{\mathcal{C}}}

\def\bmu{{\boldsymbol \mu}}

\def\bth{{\boldsymbol \theta}}

\def\bSig{{\boldsymbol \Sigma}}
\def\bGa{{\boldsymbol \Gamma}}
\def\R{R_{\mathrm{new}}}
\def\new{\mathrm{new}}

\def\s{{\bf s}}
\def\W{{\bf W}}
\newtheorem{Theorem}{\underline{\bf Theorem}}
\newtheorem{Remark}{\underline{\bf Remark}}

\newtheorem{Definition}{Definition}

\newtheorem{Condition}{Condition}

\begin{document}
\title{\bf Semi-supervised learning using copula-based regression and model averaging}

\author{Ziwen Gao$^1$, Huihang Liu$^2$ and Xinyu Zhang$^{3}$\\
	$^1$Yau Mathematical Sciences Center, Tsinghua University, Beijing, China\\
	$^2$International Institute of Finance, School of Management,\\ University of Science and Technology of China, Hefei, China\\
	$^3$Academy of Mathematics and Systems Science,\\
	Chinese Academy of Sciences, Beijing, China}
\date{}
\maketitle
\begin{abstract}
The available data in semi-supervised learning usually consists of relatively small sized labeled data and much larger sized unlabeled data.
How to effectively exploit unlabeled data is the key issue. In this paper, we write the regression function in the form of a copula and marginal distributions, and the unlabeled data can be exploited to improve the estimation of the marginal distributions.
The predictions based on different copulas are weighted, where the weights are obtained by minimizing an asymptotic unbiased estimator of the prediction risk. Error-ambiguity decomposition of the prediction risk is performed such that unlabeled data can be exploited to improve the prediction risk estimation. We demonstrate the asymptotic normality of copula parameters and regression function estimators of the candidate models under the semi-supervised framework, as well as the asymptotic optimality and weight consistency of the model averaging estimator. Our model averaging estimator achieves faster convergence rates of asymptotic optimality and weight consistency than the supervised counterpart. Extensive simulation experiments and the California housing dataset demonstrate the effectiveness of the proposed method.
\end{abstract}
{\textbf{Keywords}: asymptotic optimality, copula-based regression, model averaging, semi-supervised learning, weight consistency} 



\section{Introduction}
Semi-supervised learning has been receiving attention in both the fields of machine learning and statistics. With the improvement of data collection and storage capabilities, gathering large amounts of data has become easy. However, due to the need for human resources and expertise, the cost of labeling these data has become expensive. An example of semi-supervised learning is that hospitals usually have a large number of medical imaging images of patients, but it is tough to ask doctors to label the infected lesions in the images. Thus, semi-supervised learning usually consists of small or moderate sized labeled data and large sized unlabeled data. The analysis of semi-supervised learning becomes important when the response variables of the data are difficult or costly to obtain, and a fundamental question is how to effectively exploit unlabeled data.

Researchers in the field of machine learning primarily focus on classification problems in semi-supervised learning \citep{blum1998combining,belkin2006manifold,wang2008probability,guillaumin2010multimodal,zhang2013exploiting}.
\cite{van2020survey} provide a comprehensive review of the main approaches and recent advances in the semi-supervised classification problem in the last 20 years.
Regression problems in semi-supervised learning have also garnered significant attention \citep{wasserman2007statistical,johnson2008graph,abdel2009semi,li2017learning,timilsina2021semi}. 
\cite{kostopoulos2018semi} provide a comprehensive review of semi-supervised regression methods developed in recent years, categorizing them for analysis. 
We are concerned with how unlabeled data can be exploited to improve prediction, and the following literature is concerned with exploiting unlabeled data to improve the estimation of the coefficients of the linear model, thereby obtaining prediction. \cite{Chakrabortty2018Efficient} use a  semi-nonparametric method to impute the outcomes in unlabeled data and then fit the linear model to the unlabeled data with the imputed outcomes. 
Let the labeled data be $\calL=\{(Y_i,\X_i):i=1,\ldots,n\}$ and the unlabeled data be $\calU=\{\X_i:i=n+1,\ldots,n+N\}$,
\cite{azriel2022semi} use $\calU$ to estimate $\E(\X)$ and $\var(\X)$ in estimating the linear model coefficients in order to obtain an estimator that is better than $(\sumi\X_i\X_i\tt)^{-1}\sumi\X_iY_i$ estimated from $\calL$, where $(Y_i,\X_i)$ are observations.
\cite{song2023general} propose M-estimation in the semi-supervised framework, where the parameters are obtained by minimizing the loss on $\calL$ plus the predicted loss on $\calU$. In high-dimensional semi-supervised learning, \cite{deng2023optimal} investigate when and how to use unlabeled data to improve the estimation of linear regression parameters when linear models may be misspecified.


In addition to the aforementioned literature that establishes prediction using the linear model, there are other regression methods for predicting response variable $Y$ given covariates $\X$, such as nonparametric regression.
\cite{Hohsuk2013copula} develop the copula to obtain predictions of the regression function and enrich the family of regression estimators. Copula theory has become one of the most important methods for characterizing dependence among random variables. There are many estimation methods for copula and they are applied in finance, risk management, and other fields \citep{hu2006dependence,cai2014selection,liu2019detecting}. The reason we choose copula-based regression estimation is that copula can separate the dependence structure and marginal distributions of $(Y, \X)$. Thus, the unlabeled data $\calU$ can be exploited in estimating the marginal distributions of $\X$, which extracts information of any order moments of covariates from the unlabeled data,  achieving the goal of exploiting the unlabeled data as much as possible.
If $Y\sim Normal(\mu_y,\sigma_y^2)$ and $\X\sim Normal(\bmu_\X,\bSig_\X)$, we can use Gaussian copula to  obtain  $\mu(\x)=\mu_y+\cov(Y,\X)\bSig_\X^{-1}\left(\x-\bmu_\X\right).$ This shows that the copula-based regression can contain linear regression. In addition, the copula-based regression avoids the bandwidth selection required in nonparametric regression, and we can perform statistical inference on the correlation parameters (Theorem \ref{th:theta}) between $Y$ and $\X$, which is impossible with nonparametric regression. 
Most literature employs the copula-based regression method by selecting a copula from the candidate copulas using some model selection criteria.

As an alternative to the model selection method, model averaging provides a more robust estimator by weighting the estimators of each candidate model. Model averaging is also a widely recognized method for improving predictions, and we use it to weight candidate copulas.
Since the optimal model averaging method is proposed by \cite{hansen2007least}, there have been significant developments in the prediction problem \citep{hansen2008least,liu2020model,li2022adaboost,FANG2022219,zhang2023optimal,hu2023optimal,zhang2023model}. These studies have all demonstrated asymptotic optimality, meaning that the risk of the model averaging estimators they construct asymptotically equals the infimum of the risk over the weight space. 
However, there is currently no research on developing optimal model averaging methods under the semi-supervised learning framework, and how unlabeled data may affect asymptotic optimality. Leveraging model averaging methods in semi-supervised regression can enhance predictive performance and also enrich the applications of model averaging.

In this paper, we propose a model averaging method under the semi-supervised regression model, and we mainly exploit unlabeled data in two ways: when estimating the marginal distribution of $\X$ and when constructing the weight selection criterion. 
First, we extend the copula-based regression model proposed by \cite{Hohsuk2013copula} to the semi-supervised setting with different copulas as candidate models. Compared to \cite{Hohsuk2013copula}, we do not need to choose a copula through model selection, and we can characterize a more realistic and complex dependence structure. 
Second, according to the error-ambiguity decomposition proposed by \cite{Krogh1997advances}, we decompose the risk into two components. This allows us to exploit unlabeled data to estimate the risk when constructing the weight selection criterion. 
Finally, the predictions of each candidate model are combined using the solved weights to obtain the model averaging prediction.
Our contributions include the following four folds.
\begin{itemize}
	
	\item[(a)]{We exploit unlabeled data in each step, including estimating the copula parameters and regression functions of candidate models, as well as solving for weights. These steps utilize the information from unlabeled data as much as possible. Specifically, this is the first time that unlabeled data are exploited to estimate the distributions of covariates.}
	
	\item[(b)]{Under each candidate model, we establish the asymptotic normality of copula parameters and regression function estimators, enriching the application of copula theory in semi-supervised learning.}
	
	\item[(c)]{When all candidate models are misspecified, we obtain the asymptotic optimality based on out-of-sample risk rather than in-sample risk, making the result more practical for prediction. Meanwhile, the asymptotic optimality ensures the predictive performance of our model averaging estimator. Different from the benefits derived from unlabeled data in the existing literature, the inclusion of unlabeled data accelerates the convergence rate of asymptotic optimality.}
	
	\item[(d)]{When there are correct models among the candidate models, we prove that the  sum of the weights of these correct candidate models converges to one. In addition, the inclusion of unlabeled data accelerates the convergence rate of weights assigned to the correct candidate models.}
	
\end{itemize}

The rest of this paper is organized as follows. In Section \ref{sec:semi_modelset}, we sequentially introduce the data generation process (DGP), the construction of candidate models, and the model averaging estimator. In Section \ref{sec:semi_theory}, we present two theoretical properties of the candidate models and two theoretical properties of the model averaging estimator. The numerical results of the simulation experiments and the California housing dataset are presented in Sections \ref{sec:semi_numerical} and \ref{sec:semi_realdata}. Some concluding remarks are provided in Section \ref{sec:semi_remarks}. All the technical proofs are contained in the online Supplementary Material.

\section{Model setup}\label{sec:semi_modelset}
\subsection{Model averaging estimator}
Let $Y\in\mathbb{R}$ be the response variable
and $\X=(X_1,\ldots,X_p)\tt$ be the covariates with cumulative distribution functions $F_0$ and $F_1,\ldots,F_p$, respectively. In the semi-supervised setting, the available data are $\mathcal{D}=\calL\cup\calU$, where the labeled data $\calL=\{(Y_i,\X_i):i=1,\ldots,n\}$ are $n$ independent and identically distributed  (i.i.d.) observations and the unlabeled data $\calU=\{\X_i:i=n+1,\ldots,n+N\}$ are $N$ i.i.d. observations. Define  $\rho=n/N\in[0,+\infty)$. A common assumption is $\calL\bot\calU$, where $\bot$ represents independence. We assume that the DGP is 
\be
Y=\mu(\x)+\varepsilon, 
\ee
where $\mu(\x)=\E(Y|\X=\x)$,  $\x=(x_1,\ldots,x_p)\tt$, $\E(\varepsilon|\X)=0$, and $\var(\varepsilon)=\sigma^2<\infty$. Our goal is to use $\mathcal{D}$ to predict $\mu(\x)$ given $\X=\x$. 

In practice, regardless of which method is used to predict $\mu(\x)$, it is necessary to consider how to exploit unlabeled data. \cite{azriel2022semi} exploit unlabeled data in estimating $\E(X_j)$ and $\var(X_j)$, which extracts the information of the first and second moments of covariates from the unlabeled data. A generalized idea is to exploit unlabeled data in estimating the distribution of $X_j$, which can extract information of any order moments of covariates from the unlabeled data. Since copulas can separate the dependence structure and marginal distribution of $(Y, \X)$, we can leverage copulas to establish the estimation of the regression function $\mu(\x)$, thus allowing the use of unlabeled data in estimating the distribution of $X_j$.

We first give the definition of the copula function and then represent $\mu(\x)$ in the form of a copula function. According to Sklar’s Theorem \citep{sklar1959}, there exists a uniquely determined $p+1$ dimensional copula function $C(\cdot)$ such that for all $(y,x_1,\ldots,x_p)$,
$F_{Y,\X}(y,x_1,\ldots,x_p)=C(u_0,u_1,\ldots,u_p)$,
where $F_{Y,\X}(\cdot)$ is the joint distribution function of $(Y,\X)$, $u_0=F_0(y)$, and $u_j=F_j(x_j)$ for $j=1,\ldots,p$. 
The conditional density of $Y$ given $\X=\x$ is
$$f_0(y)\frac{c\left(u_0,u_1,\ldots,u_p\right)}{c_\x\left(u_1,\ldots,u_p\right)},$$
where $c(u_0,u_1,\ldots,u_p)=\partial^{p+1} C(u_0,u_1,\ldots,u_p)/\partial u_0\partial u_1\cdots\partial u_p$, $f_0(y)$ is the  density function of $Y$ evaluated at $y$, and  $c_\x(u_1,\ldots,u_p)=\partial^{p} C(1,u_1,\ldots,u_p)/\partial u_1\cdots\partial u_p$. 
According to  \cite{Hohsuk2013copula}, $\mu(\x)$ can be represented by the copula function:
\be\label{eq:mux_copula}
\mu(\x)=\E_Y\left\{Y\frac{c\left(F_0(Y),F_1(x_1),\ldots,F_p(x_p)\right)}{c_\x\left(F_1(x_1),\ldots,F_p(x_p)\right)}\right\}.
\ee

In \eqref{eq:mux_copula}, $c(\cdot)$ and $F_j(j=0,1,\ldots,p)$ need to be estimated. We use the rescaled empirical distribution functions to estimate $F_j$:
\be\label{hatFj}
\hat F_0(y)=\frac{1}{n+1}\sum_{i=1}^{n} \mathbb{I}(Y_i\leq y) \text{ and } \hat F_j(x_j)=\frac{1}{n+N+1}\sum_{i=1}^{n+N} \mathbb{I}(X_{ij}\leq x_j)
\ee
for $j=1,\ldots,p$, where $\mathbb{I}(\cdot)$ is the indicator function and $X_{ij}$ is the $j$th component of $\X_i$. We correct $n^{-1}$ to $(n+1)^{-1}$ in the empirical distribution function. This correction avoids the unboundedness difficulties caused when some components of the log-likelihood function of \eqref{theta_MLE} tend to 1.
In \eqref{hatFj}, we exploit the unlabeled data $\calU$ to estimate $F_j(j=1,\ldots,p)$. 

The true copula function $C(\cdot)$ is unknown, and some candidate copulas need to be constructed. Due to uncertainty in selecting an appropriate copula, model averaging or mixed copula methods can be used to address model uncertainty. The mixed copula method \citep{hu2006dependence} involves simultaneously estimating the candidate copulas parameters and weights, making it not only computationally complex but also may lead to overfitting. Therefore, we use the model averaging method to estimate  $\mu(\x)$.
Suppose we have $M$ candidate copulas, and the $m$th candidate copula is $C_m(u_0,u_1,\ldots,u_p;\bth_m)$, where $\bth_m\in\mathbb{R}^{q_m}$ is a vector of unknown copula parameters. A common estimator of $\bth_m$ is the maximum likelihood estimator $\hat\bth_m$:
\be\label{theta_MLE}
\hat\bth_m=\mathop{\argmax}\limits_{\bth_m} \sumi\log\left\{c_m\left(\hat F_0(Y_i),\hat F_1(X_{i1}),\ldots,\hat F_p(X_{ip});\bth_m\right)\right\},
\ee
where $c_m(u_0,u_1,\ldots,u_p;\bth_m)=\partial^{p+1} C_m(u_0,u_1,\ldots,u_p;\bth_m)/\partial u_0\partial u_1\cdots\partial u_p$.
In the $m$th candidate copula, $\mu(\x)$ can be estimated by
\be\label{eq:hatmum}
\hat\mu_{m}(\x)=\frac{1}{n}\sumi Y_i\frac{c_m\left(\hat F_0(Y_i),\hat F_1(x_1),\ldots,\hat F_p(x_p);\hat\bth_m\right)}{c_{\x,m}\left(\hat F_1(x_1),\ldots,\hat F_p(x_p);\hat\bth_m\right)},
\ee
where $c_{\x,m}(\hat F_1(x_1),\ldots,\hat F_p(x_p);\hat\bth_m)=n^{-1}\sumi c_m(\hat F_0(Y_i),\hat F_1(x_1),\ldots,\hat F_p(x_p);\hat\bth_m)$ according to \cite{Hohsuk2013copula}. 
Further, the model averaging estimator of $\mu(\x)$ is
\be
\hat\mu_\w(\x)=\summ w_m\hat\mu_{m}(\x),
\ee
where $\w=(w_1,\ldots,w_M)\tt$ is a weight vector belongs to the set $\calW=\{\w\in[0,1]^M: \summ w_m=1\}$.


\subsection{Weight selection criterion}
In this subsection, we propose a weight selection criterion for estimating the out-of-sample prediction risk of the model averaging estimator.

Given a new sample $(\Xn,Y_\new)$ independent of $\mathcal{D}=\calL\cup\calU$, our goal is to choose weights such that the risk function $\R(\w)=\E_{\new}\left\{\hat\mu_\w(\X_\new)-\mu(\X_\new)\right\}^2$ of the new sample is minimized, where $\E_\new$ is the expectation taken over $(\Xn,Y_\new)$ and $Y_\new=\mu(\X_\new)+\varepsilon_\new$. It is seen that $R_\new(\w)=\E_{\new}\{\hat\mu_\w(\X_\new)-Y_\new\}^2-\sigma^2$, where the first term is prediction risk and the second term is  unrelated to $\w$. Therefore, minimizing the risk function $R_\new(\w)$ with respect to $\w$ and prediction risk are equivalent, and our goal becomes minimizing the prediction risk to choose the weights.
According to the error-ambiguity decomposition proposed by \cite{Krogh1997advances}, we decompose the prediction risk into two parts: 
\be\label{eq:Rw=EA}
\E_{\new}\{\hat\mu_\w(\X_\new)-Y_\new\}^2
&=&\E_\new\left[\summ w_m\left\{\hat\mu_m(\X_\new)-Y_\new\right\}^2\right]\n\\
&\quad&-\E_\new\left[\summ w_m\left\{\hat\mu_m(\X_\new)-\hat\mu_\w(\X_\new)\right\}^2\right].
\ee
The second term in \eqref{eq:Rw=EA} does not include information about $Y_\new$ and thus can be estimated by exploiting unlabeled data $\calU$.
The purpose of decomposing the prediction risk is to exploit unlabeled data $\calU$ when estimating prediction risk, thereby improving the estimation accuracy.

We use the $K$-fold cross-validation (CV) method to estimate the two terms on the right side of Equation \eqref{eq:Rw=EA}.
Specifically, the labeled data $\calL$ and unlabeled data $\calU$ are divided into $K$ groups, respectively (for convenience, assume that $H_1 = n/K$ and $H_2=N/K$ are two integers). 
In the $k$th group, we have data $\mathcal{D}_k\equiv\calL_k\cup\calU_k=\{(\X_i,Y_i)\}_{i=(k-1)H_1+1}^{kH_1}\cup\{\X_i\}_{i=n+(k-1)H_2+1}^{n+kH_2}$ for $k=1,\ldots,K$.
Denote the CV estimator under the $m$th candidate model by
\be
\wt\bmu_m^\calL &=& \left(\wt\mu_{m,[-1]}(\X_1),\ldots,\wt\mu_{m,[-k]}(\X_{(k-1)H_1+1}),\ldots,\wt\mu_{m,[-K]}(\X_n)\right)\tt,\n\\
\wt\bmu_m^\calU&=&\left(\wt\mu_{m,[-1]}(\X_{n+1}),\ldots,\wt\mu_{m,[-k]}(\X_{n+(k-1)H_2+1}),\ldots,\wt\mu_{m,[-K]}(\X_{n+N})\right)\tt,\n
\ee
where $\wt\mu_{m,[-k]}(\x)$ is obtained with data $\mathcal{D}_k$ excluded in Equation \eqref{eq:hatmum}. Let $\wt\bmu_\w^{\calL}=\summ w_m\wt\bmu_m^{\calL}$, $\wt\bmu_\w^{\calU}=\summ w_m\wt\bmu_m^{\calU}$, $\wt\bmu_m=(\wt\bmu_m^{\calL\top},\wt\bmu_m^{\calU\top})\tt$, and $\wt\bmu_\w=(\wt\bmu_\w^{\calL\top},\wt\bmu_\w^{\calU\top})\tt$. 
The weight selection criterion is given by
\be
\C(\w)&=&\frac{1}{n}\summ w_m\left\|\wt\bmu_m^\calL-\Y\right\|^2-\frac{1}{n+N}\summ w_m\left\|\wt\bmu_m-\wt\bmu_\w\right\|^2\n\\
&=&\left(\frac{1}{n}\left\|\wt\bmu_1^\calL-\Y\right\|^2-\frac{1}{n+N}\wt\bmu_1\tt\wt\bmu_1,\ldots,\frac{1}{n}\left\|\wt\bmu_M^\calL-\Y\right\|^2-\frac{1}{n+N}\wt\bmu_M\tt\wt\bmu_M\right)\w\n\\
&\quad&+\w\tt\left\{\frac{1}{n+N}\left\|(\wt\bmu_1,\ldots,\wt\bmu_M)_{(n+N)\times M}\right\|^2\right\}\w,
\ee
where $\Y=(Y_1,\ldots,Y_n)\tt$ and $\|\mathbf{a}\|^2=\mathbf{a}\tt\mathbf{a}$ for any matrix $\mathbf{a}$. Under certain conditions, we can obtain  $$\E\{\C(\w)\}=\E[\E_{\new}\{\hat\mu_\w(\X_\new)-Y_\new\}^2](1+o(1)),$$ which provides support for the validity of $\C(\w)$ (See online Supplementary Material for the proof).

Now, let $\hat\w=\argmin_{\w\in\calW}\C(\w)$ be the weight vector, and the minimization of $\C(\w)$ with respect to $\w$ is a quadratic programming problem. The weight $\hat\w$ can be obtained instantaneously using the quadprog package in the R language. 
Finally, we obtain the model averaging estimator $\hat\mu_{\hat\w}(\x)$ of $\mu(\x)$: $\hat\mu_{\hat\w}(\x)=\summ \hat w_m\hat\mu_{m}(\x)$, where $\hat w_m$ is the $m$th component of $\hat\w$.

%
%

\section{Asymptotic properties}\label{sec:semi_theory}
In this section, we provide the asymptotic behaviors of the candidate models and the proposed model averaging estimator. Unless otherwise stated, the following limiting results are obtained by letting $n\to\infty$ and $K,p,q_1,\ldots,q_M$ are all fixed.

\subsection{The asymptotic distribution of parameter estimator}
We investigate the asymptotic distribution of $\hat\bth_m$ in each candidate model, and give some definitions first. 

\begin{Definition}\citep{shorack1972functions}.
	Let $\mathcal{Q}$ denote the set of all continuous functions $q$ on $[0,1]$, which are positive on $(0,1)$, symmetric about $0.5$, increasing on $[0,0.5]$, and for which $\int_0^1\{q(u)\}^{-2}du<\infty$.
\end{Definition}

\begin{Definition}\citep{shorack1972functions}.
	A positive function $r$ defined on $(0,1)$ is called u-shaped when it is symmetric about $0.5$ and increasing on $(0,0.5]$. When $0<\beta<1$ we define $r_\beta(u)=r(\beta u)$ for $0<u\leq 0.5$ and $r_\beta(u)=r(1-\beta(1-u))$ for $0.5<u\leq1$. There exists a constant $R_\beta$ for all $\beta>0$ in a neighbourhood of $0$, such that $r_\beta\leq R_\beta r$ on $(0,1)$, then $r$ is called a reproducing u-shaped function. We use $\mathcal{R}$ to denote the set of reproducing u-shaped functions.
\end{Definition}

\begin{Definition} 
	When $C(\cdot)$ belongs to a parametric copula family $\mathcal{C}=\{C(\cdot;\bth),\bth\in\Theta\subset\mathbb{R}^q\}$, define $\bth_0\in\Theta$ as the true copula parameter vector.
\end{Definition}

\begin{Definition} 
	Denote $\bth_m^\ast$ as the unique minimizer of
	\be
	I(\bth_m)=\int_{[0,1]^{p+1}}\log\left\{\frac{c(u_0,u_1,\ldots,u_p)}{c_m(u_0,u_1,\ldots,u_p;\bth_m)}\right\}dC(u_0,u_1,\ldots,u_p),
	\ee
	where $\bth_m^\ast$ minimizes the Kullback–Leibler (KL) loss $I(\bth_m)$ between the true model and the $m$th candidate model. 
	Define 
	\be
	\mu_m^\ast(\x)=\frac{\E\left\{Yc_m(F_0(Y),u_1,\ldots, u_p;\bth_m^\ast)\right\}}{\int_{0}^1c_{m}(u_0,u_1,\ldots,u_p;\bth_m^\ast)du_0}\n
	\ee
	as the mean regression function of the $m$th candidate model.
	When the $m$th candidate copula correctly specifies the true copula $C(\cdot)$, we have $\bth_m^\ast=\bth_{0}$ and $\mu_m^\ast(\x)=\mu(\x)$.
\end{Definition}
Given the aforementioned definitions, we list several conditions:
\begin{Condition}\label{con1}
	For each $\bth_m\in\mathbb{R}^{q_m}$, $l_{\bth_m}(\cdot;\bth_m)=\partial\log\{c_m(\cdot;\bth_m)\}/\partial\bth_m$ is continuously differentiable  on $(0,1)^{p+1}$ with $l_{\bth_m}^{(j)}(\cdot;\bth_m)=\partial l_{\bth_m}(\cdot;\bth_m)/\partial u_j$ for $j=0,1,\ldots,p$. 
	Let $l_{\bth_m,t}$ be the $t$-th element of $l_{\bth_m}(\cdot;\bth_m)$ and $l_{\bth_m,t}^{(j)}$ be the $t$-th element of $l_{\bth_m}^{(j)}(\cdot;\bth_m)$ for $t=1,\ldots,q_m$ and $j=0,1,\ldots,p$. There exist functions $r_{tj}\in\mathcal{R}$, $\wt r_{tj}\in\mathcal{R}$, and $q_{tj}\in\mathcal{Q}$, such that 
	\be
	&\left|l_{\bth_m,t}(u_0,u_1,\ldots,u_p;\bth_m)\right|\leq \prod_{j=0}^p r_{tj}(u_j), \n\\
	&\left|l_{\bth_m,t}^{(j)}(u_0,u_1,\ldots,u_p;\bth_m)\right|\leq \wt r_{tj}(u_j)\prod_{i\neq j} r_{ti}(u_i),\n\\
	&\int_{(0,1)^{p+1}}\left\{\prod_{j=0}^p r_{tj}(u_j)\right\}^2dC(u_0,u_1,\ldots,u_p)<\infty,\n
	\ee
	and
	\be
	&\int_{(0,1)^{p+1}}\left\{q_{tj}(u_j)\wt r_{tj}(u_j)\prod_{i\neq j} r_{ti}(u_i)\right\}dC(u_0,u_1,\ldots,u_p)<\infty\n
	\ee
	for $t=1,\ldots,q_m$ and $j=0,1,\ldots,p$.
\end{Condition}

\begin{Remark}
	Condition \ref{con1} is regular in deriving  the asymptotic normality of the rank statistics and can be found in \cite{ruymgaart1973asymptotic,genest1995semiparametric,tsukahara2005semiparametric}. 
	Many commonly used copulas satisfy Condition \ref{con1} \citep{tsukahara2005semiparametric}.
\end{Remark}

\begin{Condition}\label{con2}
	$l_{\bth_m}(\cdot;\bth_m)$ is differentiable with respect to $\bth_m$ and let $l_{\bth_m\bth_m}(\cdot;\bth_m)=\partial^2\log\{c_m(\cdot;\bth_m)\}/\partial\bth_m\partial\bth_m^\top$. $l_{\bth_m\bth_m}(u_0,u_1,\ldots,u_p;\bth_m)$ is continuous with respect to $u_j(j=0,\ldots,p)$ and $\bth_m$, and the elements in $l_{\bth_m\bth_m}(\cdot;\bth_m)$ are dominated by integrable function with respect to $dC(u_0,u_1,\ldots,u_p)$. 
	The $q_m\times q_m$ Fisher information matrix 
	$\bGa_m=-\E\left\{l_{\bth_m\bth_m}\left(F_0(Y), F_1(X_{1}),\ldots, F_p(X_{p});\bth_m^\ast\right)\right\}$ is nonsingular. 
\end{Condition}

\begin{Remark}
	Condition \ref{con2} involves assumptions about differentiable, continuous, and nonsingular, which are common in maximum likelihood estimation theory.
\end{Remark}

Before giving Theorem \ref{th:theta}, we define 
\be
\s(\bth_m^\ast)&=&l_{\bth_m}\left(F_0(Y), F_1(X_{1}),\ldots, F_p(X_{p});\bth_m^\ast\right),\n\\
\W_0(Y)&=&\int\left\{\mathbb{I}(F_0(Y)\leq u_0)-u_0\right\}l_{\bth_m}^{(0)}\left(u_0, u_1,\ldots, u_p;\bth_m^\ast\right)dC(u_0, u_1,\ldots, u_p),\n\\
\W_j(X_j)&=&\int\left\{\mathbb{I}(F_j(X_{j})\leq u_j)-u_j\right\}l_{\bth_m}^{(j)}\left(u_0, u_1,\ldots, u_p;\bth_m^\ast\right)dC(u_0, u_1,\ldots, u_p)\n
\ee
for $j=1,\ldots,p$.

\begin{Theorem}\label{th:theta}
	Under Conditions \ref{con1}-\ref{con2}, $n^{-1}\sumi l_{\bth_m}(\hat F_0(Y_i),\hat F_1(X_{i1}),\ldots,\hat F_p(X_{ip});\bth_m)=\mathbf{0}_{q_m}$ has a solution $\hat \bth_m$ with probability tending to one, and $\hat \bth_m$ is a consistent estimator of $\bth_m^\ast$, where $\mathbf{0}_{q_m}$ is a $q_m$-dimensional vector with all elements 0. Further, 
	\be
	\sqrt{n}(\bGa_m^{-1}\bSig_{0,m}\bGa_m^{-1})^{-1/2}(\hat\bth_m-\bth_m^\ast) \to_d \mathrm{Normal} (\mathbf{0}_{q_m},\I_{q_m}),
	\ee
	where $\I_{q_m}$ is a $q_m$-dimensional identity matrix and 
	\be
	&&\bSig_{0,m}\n\\
	&=&\var\left\{\s(\bth_m^\ast)+\W_0(Y)+\frac{\rho}{1+\rho}\sum_{j=1}^p \W_j(X_j)\right\}+\frac{\rho}{1+2\rho+\rho^2}\var\left\{\sum_{j=1}^p\W_j(X_j)\right\}.\n
	\ee
	In addition, if the $m$th candidate copula correctly specifies the true copula $C(\cdot)\in\mathcal{C}$, then $\sqrt{n}\bSig_{\mathrm{true}}^{-1/2}(\hat\bth_m-\bth_0) \to_d \mathrm{Normal} (\mathbf{0}_{q},\I_{q})$, where $\bSig_{\mathrm{true}}$ is the result of replacing $c_m(\cdot)$ and $\bth_m^\ast$ with the true copula density $c(\cdot)$ and true copula parameter $\bth_0$ in the term $\bGa_m^{-1}\bSig_{0,m}\bGa_m^{-1}$.
\end{Theorem}

Theorem \ref{th:theta} guarantees the existence and consistency of $\hat\bth_m$ and establishes the asymptotic normality of $\hat\bth_m$ of each candidate model when exploiting unlabeled data. It is the standard result of the maximum likelihood theory. See online Supplementary Material for the proof of Theorem \ref{th:theta}.

\begin{Remark}
	These terms $\W_0(Y)$ and $\W_j(X_j)$ in $\bSig_{0,m}$ are introduced in the estimation of the marginal distributions $F_0(y)$ and $F_j(x_j)(j=1,\ldots,p)$. If $F_0(y)$ and $F_j(x_j)$ are known, $\bSig_{0,m}=\var\left\{\s(\bth_m^\ast)\right\}$.
\end{Remark}

\subsection{The asymptotic distribution of $\hat\mu_m(\x)$}
We investigate the asymptotic distribution of $\hat\mu_m(\x)$ in each candidate model. 
Let $c_m^{(j)}(\cdot;\bth_m)=\partial c_m(\cdot;\bth_m)/\partial u_j$ for $j=0,1,\ldots,p$, $\mathbf{\dot c}_{m}(\cdot;\bth_m)=\partial c_m(\cdot;\bth_m)/\partial\bth_m$,
\be
&&{E}_{i0}(\bth_m^\ast)\n\\
&=&-\int\left\{\mathbb{I}(Y_i\leq y)-F_0(y)\right\}c_m(F_0(y),u_1,\ldots,u_p;\bth_m^\ast)dy\n\\
&\quad&+\left\{l_{\bth_m}\left(F_0(Y_i), F_1(X_{i1}),\ldots, F_p(X_{ip});\bth_m^\ast\right)\right\}\tt\E\left\{Y \mathbf{\dot c}_{m}(F_0(Y),u_{1},\ldots, u_{p};\bth_{m}^\ast)\right\}\n\\
&\quad&+\left\{\W_0(Y_i)\right\}\tt\E\left\{Y \mathbf{\dot c}_{m}(F_0(Y),u_{1},\ldots, u_{p};\bth_{m}^\ast)\right\}\n
\ee
for $i=1,\ldots,n$,  
\be
{E}_{ij}(\bth_m^\ast)&=&\sum_{j=1}^{p}\left\{\W_j(X_{ij})\right\}\tt\E\left\{Y \mathbf{\dot c}_{m}(F_0(Y),u_{1},\ldots, u_{p};\bth_{m}^\ast)\right\}\n\\
&\quad&+\left\{\mathbb{I}(X_{ij}\leq x_{j})-F_j(x_j)\right\}\E\left\{Y c_m^{(j)}(F_0(Y),u_{1},\ldots, u_{p};\bth_{m}^\ast)\right\}\n
\ee
for $i=1,\ldots,n+N$ and $j=1,\ldots,p$, and 
\be
\sigma_0^2(\bth_m^\ast)=\var\left(E_{10}(\bth_m^\ast)+\frac{\rho}{1+\rho}\sum_{j=1}^pE_{1j}(\bth_m^\ast)\right)+\frac{\rho}{1+2\rho+\rho^2}\var\left(\sum_{j=1}^pE_{1j}(\bth_m^\ast)\right).\n
\ee
\begin{Condition}\label{cmj_conti}
	For $j=0,1,\ldots,p$, $c_m^{(j)}(u_0,u_1,\ldots,u_p;\bth_m)$ and $\mathbf{\dot c}_{m}(u_0,u_1,\ldots,u_p;\bth_m)$ are continuous at $(u_1,\ldots,u_p,\bth_m)$ uniformly for $u_0$. In addition, $c_m^{(j)}(u_0,u_1,\ldots,u_p;\bth_m)$ and $\mathbf{\dot c}_{m}(u_0,u_1,\ldots,u_p;\bth_m)$ are continuous with respect to $u_0$.
\end{Condition}
\begin{Remark}
	Condition \ref{cmj_conti} specifies the continuity of some functions, and is the same as Assumption C of \cite{Hohsuk2013copula}.
\end{Remark}
\begin{Theorem}\label{th:mu}
	If $\E(Y^2)<\infty$, $\sigma_0^2(\bth_m^\ast)<\infty$, and Conditions \ref{con1}-\ref{cmj_conti} hold, we have 
	\be
	\sqrt{n}\sigma^{-1}_m(\bth_m^\ast)\left\{\hat\mu_m(\x)-\mu_m^\ast(\x)\right\}\to_d\mathrm{Normal}\left(0,1\right),
	\ee
	where 
	$\sigma^{-1}_m(\bth_m^\ast)=\{\int_{0}^1c_{m}(u_0,u_1,\ldots,u_p;\bth_m^\ast)du_0\}\sigma^{-1}_0(\bth_m^\ast)$.
	
	In addition, if the $m$th candidate copula correctly specifies the true copula $C(\cdot)\in\mathcal{C}$, then $\sqrt{n}\sigma^{-1}_{\mathrm{true}}\left\{\hat\mu_m(\x)-\mu(\x)\right\}\to_d\mathrm{Normal}\left(0,1\right)$, where $\sigma^{-1}_{\mathrm{true}}$ is the result of replacing $c_m(\cdot)$ and $\bth_m^\ast$ with the true copula density $c(\cdot)$ and true copula parameter $\bth_0$ in the term $\sigma^{-1}_m(\bth_m^\ast)$.
\end{Theorem}
Theorem \ref{th:mu} further establishes the asymptotic normality of $\hat\mu_m(\x)$ based on Theorem \ref{th:theta} when exploiting unlabeled data. See online Supplementary Material for the proof of Theorem \ref{th:mu}.

\subsection{Asymptotic optimality}
Theorems \ref{th:theta} and \ref{th:mu} are established for some estimators of each candidate model, and below we consider the asymptotic property of the model averaging estimator. We allow $M$ to tend to infinity with $n$.

We use $R^\ast(\w)=\E_\new\{\hat\mu_\w^\ast(\X_\new)-\mu(\X_\new)\}^2$ to denote the risk function calculated based on $\hat\mu_\w^\ast(\X_\new)=\summ w_m\hat\mu_m^\ast(\X_\new)$ instead of $\hat\mu_\w(\X_\new)$, where
$$\hat\mu_m^\ast(\x)=\frac{1}{n}\sumi Y_i\frac{c_m\left(\hat F_0(Y_i),\wt F_1(x_1),\ldots,\wt F_p(x_p);\bth_m^\ast\right)}{c_{\x,m}\left(\wt F_1(x_1),\ldots,\wt F_p(x_p);\bth_m^\ast\right)}$$ corresponds to \eqref{eq:hatmum}, except that $\hat\bth_m$ is replaced by $\bth_m^\ast$ and $\hat F_j(x_j)$ is replaced by $\wt F_j(x_j)=(n+1)^{-1}\sum_{i=1}^{n} \mathbb{I}(X_{ij}\leq x_j)$.
Let $\xi_n=\inf_{\w\in\calW}R^\ast(\w)$ denote the minimum risk function of the model averaging estimators using the limiting value $\bth_m^\ast$ and $a\vee b$ denote $\max\{a,b\}$.
\begin{Condition}\label{theta_var}
	$\lambda_{\mathrm{min}}\{(\bGa_m^{-1}\bSig_{0,m}\bGa_m^{-1})^{-1}\}\geq c_{\mathrm{min}}$ almost surely for $m=1,\ldots,M$, where $\lambda_{\mathrm{min}}(\cdot)$ is the minimum eigenvalue of a matrix and $c_{\mathrm{min}}$ is a positive constant.
\end{Condition}
\begin{Remark}
	According to Theorem \ref{th:theta}, we know that $\bGa_m^{-1}\bSig_{0,m}\bGa_m^{-1}$ is a positive definite matrix for any $m$, and Condition \ref{theta_var} is mild. Actually, Condition \ref{theta_var} is the primitive assumption of Assumption 1 in \cite{zhang2023model}, and  Condition \ref{theta_var} is used to obtain $\|\hat\bth_m-\bth_m^\ast\|=O_p(M^{1/2}n^{-1/2})$ uniformly for $m=1,\ldots,M$. See  online Supplementary Material.
\end{Remark}

\begin{Condition}\label{mu_bounded}
	$\E_\new\{\hat\mu_m^\ast(\X_\new)\}^2$ and $\mu_m^\ast(\x)$ are bounded uniformly for $m=1,\ldots,M$ and $\x\in\mathbb{R}^p$.
\end{Condition}
\begin{Remark}
	Condition \ref{mu_bounded} requires that $\E_\new\{\hat\mu_m^\ast(\X_\new)\}^2$ and $\mu_m^\ast(\x)$ are uniformly  bounded.
	Similar conditions can be found in Assumption 2 of \cite{zhang2023model} and Condition 4 of \cite{yu2018asymptotic}.
\end{Remark}
\begin{Condition}\label{con:CLT}
	$\var\left\{\mu_{m}^\ast(\X_i)-Y_i\right\}$ and $\var\left[\left\{\mu_m^\ast(\X_i)-\mu_{m_1}^\ast(\X_i)\right\}\left\{\mu_m^\ast(\X_i)-\mu_{m_2}^\ast(\X_i)\right\}\right]$ are bounded uniformly for $m=1,\ldots,M$, $m_1=1,\ldots,M$, and $m_2=1,\ldots,M$.
\end{Condition}
\begin{Remark}
	Condition \ref{con:CLT} requires that the variances of  $\left\{\mu_{m}^\ast(\X_i)-Y_i\right\}$ and $\{\mu_m^\ast(\X_i)-\mu_{m_1}^\ast(\X_i)\}\{\mu_m^\ast(\X_i)-\mu_{m_2}^\ast(\X_i)\}$ exist uniformly for $m,m_1,m_2$, and this condition is used to apply the central limit theorem. Similar conditions can be found in Assumption 4 of \cite{zhang2023model}, Assumption 5 of \cite{zhang2023optimal}, and Condition 3 of \cite{gao2023reliability}.
\end{Remark}

\begin{Condition}\label{xi_n}
	$\left[\left(M^{1/2}n^{-1/2}\right)\vee \left\{M^{3/2}(n+N)^{-1/2}\right\}\right]\xi_n^{-1}=o_p(1)$.
\end{Condition}
\begin{Remark}
	Condition \ref{xi_n} requires that all candidate models be misspecified. If the $m^\ast$th candidate model is correctly specified, we have $\bth_{m^\ast}^\ast=\bth_{0}$, $\mu_{m^\ast}^\ast(\x)=\mu(\x)$, and  $$\xi_n=\inf_{\w\in\calW}R^\ast(\w)\leq R^\ast(\w_{m^\ast})=\E_\new\left\{\hat\mu_{m^\ast}^\ast(\X_\new)-\mu(\X_\new)\right\}^2=O_p(n^{-1}),$$ where $\w_{m^\ast}$ is the weight vector whose the $m^\ast$th component is one and the others are zeros, and the last step is similar to the proof of Theorem \ref{th:mu}. Thus, Condition \ref{xi_n} is violated. Condition \ref{xi_n} is widely used in the literature of model averaging, and is similar to Condition (8) of \cite{wan2010least}, Assumption 2.3 of \cite{liu2013heteroscedasticity},  Condition C4 of \cite{fang2019model}, and Assumption 5 of  \cite{zhang2024prediction}, etc.
	In the next subsection, we give the asymptotic property when the candidate model set contains the correct models.
	
	Condition \ref{xi_n} becomes $M^{3/2}n^{-1/2}\xi_n^{-1}=o_p(1)$ when $N=0$ and $M^{1/2}n^{-1/2}\xi_n^{-1}=o_p(1)$ when $N\geq M^2n-n$. It can be seen that Condition \ref{xi_n} when $N=0$ is stronger than Condition \ref{xi_n} when $N\geq M^2n-n$.
\end{Remark}

\begin{Theorem}\label{th:opt}
	If Conditions \ref{con1}-\ref{xi_n} and $\E(Y^2)<\infty$ hold, we have 
	\be
	\frac{R_\new(\hat\w)}{\inf_{\w\in\calW}R_\new(\w)}=1+O_p\left(\sqrt{\frac{M}{n\xi_n^2}}\vee \sqrt{\frac{M^3}{(n+N)\xi_n^2}}\right)=1+o_p(1).
	\ee
\end{Theorem}
Theorem \ref{th:opt} establishes the asymptotic optimality of our model averaging estimator in the sense that the out-of-sample risk of $\hat\mu_{\hat\w}(\x)$ is asymptotically identical to that of the infeasible best model averaging estimator. Asymptotic optimality provides guarantees for our method in terms of prediction, and the inclusion of unlabeled data can improve the convergence rate of asymptotic optimality and weaken Condition \ref{xi_n}.
See online Supplementary Material for the proof of Theorem \ref{th:opt}.

\begin{Remark}
	From Theorem \ref{th:opt}, it follows that the rate of convergence of asymptotic optimality is related to $\xi_n$, $M$, $n$, and $N$. The rate of convergence of asymptotic optimality is $O_p(M^{3/2}n^{-1/2}\xi_n^{-1})$ when there are no unlabeled data ($N=0$). The rate of convergence becomes $O_p(M^{1/2}n^{-1/2}\xi_n^{-1})$ when $N\geq M^2n-n$.  
	Thus, increasing $N$ leads to the  convergence rate of asymptotic optimality increasing from $O_p(M^{3/2}n^{-1/2}\xi_n^{-1})$ to $O_p(M^{1/2}n^{-1/2}\xi_n^{-1})$.
\end{Remark}

\subsection{Convergence of weights}
The second asymptotic property of the model averaging estimator is the consistency of weights when the correct models exist in the candidate model set. If the $m$th candidate copula correctly specifies the true copula $C(\cdot)\in\mathcal{C}$, then $\bth_m^\ast=\bth_{0}$ and $\mu_m^\ast(\x)=\mu(\x)$. 

Let $\mathcal{M}$ be a subset of $\{1,\ldots,M\}$ that consists of the indices of the correctly specified models, and let $\hat\w_\triangle=\sum_{m\in\mathcal{M}}\hat w_m$ be the sum of the weights assigned to the correctly specified models.
Denote $\calW_s=\{\w\in\calW: \sum_{m\notin\mathcal{M}} w_m=1\}$ and $\zeta_n=\inf_{\w\in\calW_s}R^\ast(\w)$, we need the following further condition.
\begin{Condition}\label{con:missxi}
	$\left[\left(M^{1/2}n^{-1/2}\right)\vee \left\{M^{3/2}(n+N)^{-1/2}\right\}\right]\zeta_n^{-1}=o_p(1)$.
\end{Condition}
\begin{Remark}
	In fact, when $\mathcal{M}=\emptyset$, Condition \ref{con:missxi} is equivalent to Condition \ref{xi_n}. Thus, Condition \ref{con:missxi} is a special case of Condition \ref{xi_n}. Similarly, Condition \ref{con:missxi} when $N=0$ is stronger than Condition \ref{con:missxi} when $N\geq M^2n-n$.
\end{Remark}

\begin{Theorem}\label{th:weight}
	If Conditions \ref{con1}-\ref{con:CLT}, and \ref{con:missxi} are satisfied, then 
	\be
	(1-\hat\w_\triangle)^2=O_p\left(\sqrt{\frac{M}{n\zeta_n^2}}\vee \sqrt{\frac{M^3}{(n+N)\zeta_n^2}}\right)=o_p(1)
	\ee
	and $\hat\w_\triangle\to1$ in probability.
\end{Theorem}
Theorem \ref{th:weight} shows that when the candidate model set includes the correct models, the weights are consistently concentrated on these correct models.
See online Supplementary Material for the proof of Theorem \ref{th:weight}. 
\begin{Remark}
	Similar to the effect of $N$ on the convergence rate of the asymptotic optimality, the increase of $N$ also accelerates the convergence rate of $\hat\w_\triangle$ to 1 and weakens Condition \ref{con:missxi}.
\end{Remark}

\section{Simulation studies}\label{sec:semi_numerical}
In this section, the five DGPs designed by \cite{azriel2022semi} are used to evaluate the performance of the proposed model averaging estimator and verify Theorems \ref{th:opt} and \ref{th:weight}. We also assess the performance of our method under a simulated data version of the Los Angeles homeless dataset provided by \cite{song2023general}.

\subsection{Comparison of mean squared prediction error via simulation}\label{sec:simulation}
This subsection compares our model averaging estimator with six semi-supervised methods through the five DGPs designed by \cite{azriel2022semi}. The specific methods are described as follows, and all methods utilize the whole available data $\mathcal{D}$.
\begin{itemize}
	\item[(a)]{Our model averaging estimator ($K$-CRMA). Following \cite{breiman1992submodel} and \cite{hastie2001elements}, we consider $K=\{5,10\}$.
	}
	
	\item[(b)]{The projection-based semi-supervised estimator (PSSE) method proposed by \cite{song2023general}.}
	
	\item[(c)]{The partial-information-based semi-supervised estimator (PI) proposed by \cite{azriel2022semi}.}
	
	\item[(d)]{The efficient and adaptive semi-supervised estimator (EASE) proposed by  \cite{Chakrabortty2018Efficient}.}
	
	\item[(e)]{
		Model averaging estimator with $w_m={\exp(-\mathrm{BIC}_m/2)}/ {\sum_{i=1}^M\exp(-\mathrm{BIC}_i/2)}$
		$(m=1,\ldots,M)$, where $\mathrm{BIC}_m =-2l_n(\hat\bth_m)+\log(n)q_m$, $l_n(\hat\bth_m)$ is the loglikelihood function
		evaluated at the estimator $\hat\bth_m$, and
		$q_m$ is the dimension of $\hat\bth_m$. This weight form, proposed by  \cite{buckland1997model}, we call  it ``SBIC" and this SBIC method is an approximation of Bayesian model averaging.}
	
	\item[(f)]{Model selection estimator with $w_b=1$ and $b=\argmin_{m} \mathrm{BIC}_m$. This is the BIC model selection method \citep{schwarz1978estimating}, and we call it ``BICMS''.}
	
	\item[(g)]{Model averaging estimator with $\w=(1/M,\ldots,1/M)\tt$. This form shows each candidate copula is assigned equal weight, and we call  it  ``EWMA''.}
\end{itemize}

In addition to the methods mentioned above, we also compare our method, named ``5-LABEL", which only exploits labeled data and $K=5$.
We set the dimension of the covariates $p=\{4,7\}$, the number of labeled data $n=\{200,500\}$, and the number of unlabeled data $N=\{n,2n,4n\}$. These settings give rise to 12 combinations of $(p,n,N)$, and all methods are repeated $500$ times under each combination. 
Five DGPs in \cite{azriel2022semi} are considered, and the distributions of $\X$ and $\varepsilon$ in all DGPs are $\mathrm{Normal}_p(\mathbf{0},\I_p)$ and $\mathrm{Normal}(0,2)$, respectively, where $\I_p$ is a $p$-dimensional identity matrix. 
\begin{itemize}
	\item[(a)] DGP1: $Y=\mathbf{1}_p\tt\X+\varepsilon$, where $\mathbf{1}_p$ is a $p$-dimensional vector with all elements 1. 
	
	\item[(b)] DGP2: $Y=\mathbf{1}_p\tt\X+\mathbf{1}_p\tt\X^2+\varepsilon$, where $\X^2=(X_1^2,\ldots,X_p^2)\tt$.
	
	\item[(c)] DGP3: $Y=\mathbf{1}_p\tt\X+0.3\times\mathbf{1}_p\tt\X^2+\varepsilon$.
	
	\item[(d)] DGP4: $Y=\mathbf{1}_p\tt\X+\mathbf{1}_p\tt\left\{\X^3-\X^2+\exp(\X)\right\}+\varepsilon$, where $\X^3=(X_1^3,\ldots,X_p^3)\tt$ and $\exp(\X)=(\exp(X_1),\ldots,\exp(X_p))\tt$.
	
	\item[(e)] DGP5: $Y=\mathbf{1}_p\tt\X+0.3\times\mathbf{1}_p\tt\left\{\X^3-\X^2+\exp(\X)\right\}+\varepsilon$.
\end{itemize}

A lot of literature \citep{Hohsuk2013copula,cai2014selection,liu2019detecting} recommends using the Gaussian, Student's $t$, Gumbel, Clayton, Frank, and Joe copulas. We set our first six candidate copulas as these six individual copulas, and the last candidate copula as a mixture of these six copulas \citep{liu2019detecting,gao2023reliability}, so $M=7$. We focus on the out-of-sample prediction performance and report the mean squared prediction error (MSPE) for all methods:  MSPE=$(n+N)^{-1}\sum_{i=1}^{n+N}(Y_{i,\new}-\hat Y_{i,\new})^2$, where $\{Y_{i,\new}\}_{i=1}^{n+N}$ are new observations independent of $\mathcal{D}$ and $\hat Y_{i,\new}$ is the prediction of $Y_{i,\new}$ obtained using a certain method.
The mean values of 500 simulation results for each combination of $(p,n,N)$ under five DGPs are displayed in Tables \ref{tab:AzrielDGP_p=4}-\ref{tab:AzrielDGP_p=7}.

The main findings in Tables \ref{tab:AzrielDGP_p=4}-\ref{tab:AzrielDGP_p=7} are summarized as follows: the MSPE of the 5-CRMA method is always lower than that of the 5-LABEL method, indicating that the exploitation of unlabeled data enhances the predictive performance.
The 5/10-CRMA method performs best in most cases for DGP2-DGP5 and is close to the best method in other cases.
In addition, the change of $K$ has a minor impact on the performance of our proposed method.
The MSPEs for the PSSE, PI, and EASE methods are almost identical, and these three methods perform best in DGP1.	
The BICMS and SBIC methods have the smallest MSPE under DGP4 when $p = 4$.  The SBIC method often produces a smaller MSPE than the BICMS method.
The EWMA method always performs poorly, an indication that simple weighting does not work and some strategies should be applied during weight selection.
In summary, the $K$-CRMA method exhibits a clear competitive advantage among all methods.

\begin{table}[htbp]
	\setlength{\tabcolsep}{1mm}
	\centering
	\caption{Comparison of MSPEs for different methods when $p=4$.}\label{tab:AzrielDGP_p=4}
	\begin{tabular}{ccrrcrrrrrr}
		\hline
		$(n,N)$ &    & \multicolumn{1}{c}{5-CRMA} & \multicolumn{1}{c}{10-CRMA} &5-LABEL & \multicolumn{1}{c}{PSSE} & \multicolumn{1}{c}{PI} & \multicolumn{1}{c}{EASE} & \multicolumn{1}{c}{BICMS} & \multicolumn{1}{c}{SBIC} & \multicolumn{1}{c}{EWMA} \\
		\hline
		\multirow{5}[2]{*}{(200,200)} & DGP1  & 4.16  & 4.16 & 4.19 & \textbf{4.10} & \textbf{4.10} & \textbf{4.10} & \underline{4.14}  & \underline{4.14}  & 5.45 \\
		& DGP2  & \textbf{10.95} & \textbf{10.95} & \underline{11.01}   & 12.33 & 12.39 & 12.60 & 13.09 & 13.10 & 13.48 \\
		& DGP3  & \underline{4.85}  & \textbf{4.84}  & 4.90 & \textbf{4.84} & \textbf{4.84} & \underline{4.85}  & 4.90  & 4.90  & 6.28 \\
		& DGP4  & 28.14 & \underline{28.12}  & 33.87  & 48.04 & 45.21 & 45.79 & \textbf{27.97} & \textbf{27.97} & 83.52 \\
		& DGP5  & \underline{6.42} & \textbf{6.41} & 7.02 & 7.79  & 7.82  & 7.86  & 7.10  & 7.07  & 15.38 \\
		\hline
		\multirow{5}[2]{*}{(200,400)} & DGP1  & 4.14  & 4.13  & 4.17 & \underline{4.09}  & \underline{4.09}  & \textbf{4.08} & 4.12  & 4.12  & 5.45 \\
		& DGP2  & \textbf{10.93} & \textbf{10.93} & \underline{11.01} & {12.30} & 12.37 & 12.59 & 13.11 & 13.09 & 13.45 \\
		& DGP3  & \textbf{4.82} & \textbf{4.82} & 4.89 & \underline{4.84}  & \underline{4.84}  & \underline{4.84}  & 4.86  & 4.86  & 6.27 \\
		& DGP4  & 25.96 & \underline{25.93} &33.13  & 44.37 & 44.75 & 45.40 & \textbf{25.81} & \textbf{25.81} & 83.37 \\
		& DGP5  & \textbf{6.15}  & \textbf{6.15} & \underline{6.92} & 7.71  & 7.75  & 7.80  & 7.00  & {6.95}  & 15.33 \\
		\hline
		\multirow{5}[2]{*}{(200,800)} & DGP1  & {4.14} & 4.14  & 4.18 & \textbf{4.10} & \textbf{4.10}  & \textbf{4.10} & \underline{4.13}  & \underline{4.13}  & 5.45  \\
		& DGP2  & \underline{10.93} & \textbf{10.92} &11.02  & 12.36 & 12.38 & 12.61 & 13.03 & {13.00} & 13.45  \\
		& DGP3  & \underline{4.82}  & \textbf{4.81} & 4.89 & 4.84  & 4.84  & 4.85  & 4.84  & 4.84  & 6.26 \\
		& DGP4  & {24.95} & \underline{24.92} & 34.01 & 45.64 & 45.76 & 46.61 & \textbf{24.77} & \textbf{24.77} & 84.55 \\
		& DGP5  & \textbf{6.04}  & \textbf{6.04} & \underline{7.01} & 7.81  & 7.85  & 7.92  & 7.12  & {7.07} & 15.43 \\
		\hline
		\multirow{5}[2]{*}{(500,500)} & DGP1  & 4.06  & 4.06  & 4.07 & \textbf{4.03} & \textbf{4.03} & \textbf{4.03} & \underline{4.05}  & \underline{4.05}  & 5.40 \\
		& DGP2  & \textbf{10.72} & \textbf{10.72} & \underline{10.77} & 12.12 & 12.15 & 12.28 & {11.83} & {11.83} & 13.36 \\
		& DGP3  & \textbf{4.73} & \textbf{4.73} & 4.76 & 4.76  & 4.76  & 4.77  & \underline{4.75}  & \underline{4.75}  & 6.22 \\
		& DGP4  & 22.72 & \underline{22.71} & 26.44 & 46.30 & 44.75 & 45.00 & \textbf{22.67} & \textbf{22.67} & 82.49 \\
		& DGP5  & \textbf{5.61} & \textbf{5.61} & \underline{5.96} & 7.68  & 7.70  & 7.71  & 6.27  & {6.24}  & 15.13 \\
		\hline
		\multirow{5}[2]{*}{(500,1000)} & DGP1  & \underline{4.07}  & \underline{4.07}  & 4.08 & \textbf{4.05} & \textbf{4.05} & \textbf{4.05} & \underline{4.07}  & \underline{4.07}  & 5.43 \\
		& DGP2  & \textbf{10.72} & \textbf{10.72} & \underline{10.79} & 12.11 & 12.15 & 12.28 & {11.80} & {11.80} & 13.35 \\
		& DGP3  & \textbf{4.73} & \textbf{4.73} & 4.77 & 4.78  & 4.78  & 4.78  & \underline{4.75}  & \underline{4.75}  & 6.23 \\
		& DGP4  & \underline{21.25} & \textbf{21.24} & 26.23 & 43.85 & 44.41 & 44.71 & \textbf{21.24} & \textbf{21.24} & 82.39 \\
		& DGP5  & \textbf{5.49} & \textbf{5.49} & \underline{5.96} & 7.65  & 7.67  & 7.70  & 6.24  & {6.22}  & 15.15 \\
		\hline
		\multirow{5}[2]{*}{(500,2000)} & DGP1  & \underline{4.06}  & \underline{4.06}  & 4.07 & \textbf{4.04}  & \textbf{4.04}  & \textbf{4.04} & \underline{4.06}  & \underline{4.06}  & 5.43  \\
		& DGP2  & \textbf{10.69} & \textbf{10.69} & \underline{10.79} & 12.11 & 12.16 & 12.29 & {11.77} & {11.77} & 13.36  \\
		& DGP3  & \textbf{4.72} & \textbf{4.72} & 4.77 & 4.77  & 4.78  & 4.78  & \underline{4.73}  & \underline{4.73}  & 6.24  \\
		& DGP4  & \underline{20.01} & \underline{20.01} & 26.08 & 49.84 & 44.24 & 44.55 & \textbf{20.00} & \textbf{20.00} & 82.13  \\
		& DGP5  & \textbf{5.37} & \textbf{5.37} & \underline{5.94} & 7.63  & 7.66  & 7.69  & 6.25  & {6.22}  & 15.13  \\\hline
	\end{tabular}%
	\begin{tablenotes}
		\footnotesize
		\item[1] Note: The smallest MSPE in each row is shown in bold, and the second smallest MSPE is underlined.
	\end{tablenotes}
\end{table}%

\begin{table}[htbp]
	\setlength{\tabcolsep}{1mm}
	\centering
	\caption{Comparison of MSPEs for different methods when $p=7$.}\label{tab:AzrielDGP_p=7}
	\begin{tabular}{ccrrcrrrrrr}
		\hline
		$(n,N)$ &    & \multicolumn{1}{c}{5-CRMA} & \multicolumn{1}{c}{10-CRMA} &5-LABEL & \multicolumn{1}{c}{PSSE} & \multicolumn{1}{c}{PI} & \multicolumn{1}{c}{EASE} & \multicolumn{1}{c}{BICMS} & \multicolumn{1}{c}{SBIC} & \multicolumn{1}{c}{EWMA} \\
		\hline
		\multirow{5}[0]{*}{(200,200)} & DGP1  & 4.26  & 4.26 & 4.32  & \textbf{4.15} & \textbf{4.15} & \textbf{4.15} & \underline{4.24}  & \underline{4.24}  & 7.01 \\
		& DGP2  & \underline{16.86} & \textbf{16.85} &  17.07 & 18.84 & 18.90 & 19.04 & 24.25 & 24.37 & 21.11 \\
		& DGP3  & 5.61  & 5.60  & 5.77  & \textbf{5.49} & \textbf{5.49} & \underline{5.50}  & 5.79  & 5.78  & 8.42 \\
		& DGP4  & \underline{50.81} & \textbf{50.61} & 62.99  & 76.32 & 76.96 & 77.64 & 52.81 & 52.60 & 158.64 \\
		& DGP5  & \underline{8.55} & \textbf{8.53}  & 10.05  & 10.70 & 10.73 & 10.80 & 10.93 & 10.88 & 26.38 \\\hline
		\multirow{5}[0]{*}{(200,400)} & DGP1  & 4.24  & 4.24  & 4.31  & \textbf{4.15} & \underline{4.16}  & \textbf{4.15} & 4.22  & 4.23  & 7.03 \\
		& DGP2  & \underline{16.88} & \textbf{16.87} & 17.18  & 18.78 & 18.90 & 19.04 & 24.33 & 24.46 & 21.15 \\
		& DGP3  & 5.55  & \underline{5.54}  & 5.77  & \textbf{5.47} & \textbf{5.47} & \textbf{5.47} & 5.73  & 5.71  & 8.42 \\
		& DGP4  & \underline{46.23} & \textbf{46.14} & 63.68  & 76.95 & 77.82 & 78.63 & 49.18 & 48.85 & 160.35 \\
		& DGP5  & \underline{7.93} & \textbf{7.92} & 10.04  & 10.75 & 10.77 & 10.84 & 10.94 & 10.91 & 26.54 \\\hline
		\multirow{5}[0]{*}{(200,800)} & DGP1  & 4.27  & 4.27  &4.35   & \textbf{4.18}  & \underline{4.19}  & \textbf{4.18}  & 4.25  & 4.25  & 7.07  \\
		& DGP2 & \underline{16.75} & \textbf{16.74} & 17.09   & 18.73 & 18.87 & 19.05 & 24.25 & 24.36 & 21.04 \\
		& DGP3  & 5.52  & 5.52  & 5.79  & \textbf{5.49}  & \underline{5.50}  & \underline{5.50}  & 5.66  & 5.66  & 8.43  \\
		& DGP4  & \underline{41.70} & \textbf{41.66} & 63.69  & 76.20 & 77.27 & 78.11 & 45.05 & 45.01 & 161.41 \\
		& DGP5  & \underline{7.44}  & \textbf{7.43}  &  10.12 & 10.74 & 10.78 & 10.85 & 10.97 & 10.94 & 26.73  \\\hline
		\multirow{5}[0]{*}{(500,500)} & DGP1  & 4.11  & 4.11  &4.13   & \textbf{4.07} & \textbf{4.07} & \textbf{4.07} & \underline{4.10}  & \underline{4.10}  & 6.98 \\
		& DGP2  & \textbf{16.34} & \textbf{16.34} & \underline{16.51}  & {18.26} & 18.32 & 18.43 & 20.54 & 20.40 & 20.93 \\
		& DGP3  & \textbf{5.34} & \textbf{5.34} & 5.42  & \textbf{5.34} & \underline{5.35}  & \underline{5.35}  & \textbf{5.34} & \textbf{5.34} & 8.34 \\
		& DGP4  & 40.01 & 39.93 & 48.60  & 74.41 & 74.75 & 75.05 & \textbf{39.85} & \underline{39.86} & 158.31 \\
		& DGP5  & \underline{7.06} & \textbf{7.05} & 8.07  & 10.40 & 10.43 & 10.46 & 10.48 & 10.47 & 26.21 \\\hline
		\multirow{5}[0]{*}{(500,1000)} & DGP1  & 4.11  & 4.11 &  4.13  & \textbf{4.07} & \textbf{4.07} & \textbf{4.07} & \underline{4.10}  & \underline{4.10}  & 6.99 \\
		& DGP2  & \textbf{16.32} & \textbf{16.32} & \underline{16.55}  & {18.30} & 18.38 & 18.50 & 20.38 & 20.32 & 20.93 \\
		& DGP3  & \textbf{5.32}  & \textbf{5.32} & 5.43  & 5.36  & 5.36  & 5.36  & \underline{5.33}  & \underline{5.33}  & 8.34 \\
		& DGP4  & {37.27} & \textbf{37.22} & 49.15  & 74.75 & 75.19 & 75.59 & \underline{37.23} & \textbf{37.22} & 159.36 \\
		& DGP5  & \underline{6.72} & \textbf{6.71} & 8.14  & 10.45 & 10.48 & 10.52 & 10.53 & 10.54 & 26.33 \\\hline
		\multirow{5}[0]{*}{(500,2000)} & DGP1  & \underline{4.09}  & \underline{4.09}  & 4.12  & \textbf{4.06}  & \textbf{4.06}  & \textbf{4.06}  & \underline{4.09}  & \underline{4.09}  & 6.97 \\
		& DGP2 & \textbf{16.26} & \textbf{16.26} & \underline{16.52}  & {18.23} & 18.32 & 18.45 & 20.37 & 20.30 & 20.87  \\
		& DGP3  & \textbf{5.28}  & \textbf{5.28}  & 5.41  & 5.34  & 5.34  & 5.34  & \underline{5.29}  & \underline{5.29}  & 8.31  \\
		& DGP4  & \underline{34.61} & \textbf{34.60} & 48.86  & 74.31 & 74.86 & 75.27 & 34.75 & 34.71 & 159.24  \\
		& DGP5  & \textbf{6.37}  & \textbf{6.37}  &\underline{8.09}  & {10.39} & 10.43 & 10.47 & 10.51 & 10.51 & 26.27 \\\hline
	\end{tabular}%
	\begin{tablenotes}
		\footnotesize
		\item[1] Note: The smallest MSPE in each row is shown in bold, and the second smallest MSPE is underlined.
	\end{tablenotes}
\end{table}%

\subsection{Verification of Theorems \ref{th:opt} and \ref{th:weight} via simulation}
In this subsection, we verify Theorems \ref{th:opt} and \ref{th:weight} using the five DGPs designed by \cite{azriel2022semi}, with the same candidate copulas in Subsection \ref{sec:simulation}. DGP1 is a linear model and both $\X$ and $\varepsilon$ follow the normal distributions, so DGP1 represents the case where the candidate models include the correct models, which can be used to verify Theorem \ref{th:weight}. DGP2 to DGP5 are cases where all candidate models are misspecified, which can be used to verify Theorem \ref{th:opt}.

\begin{figure}[htbp]
	\centering
	\subfigure[Asymptotic optimality verification ]{
		\includegraphics[width=7cm]{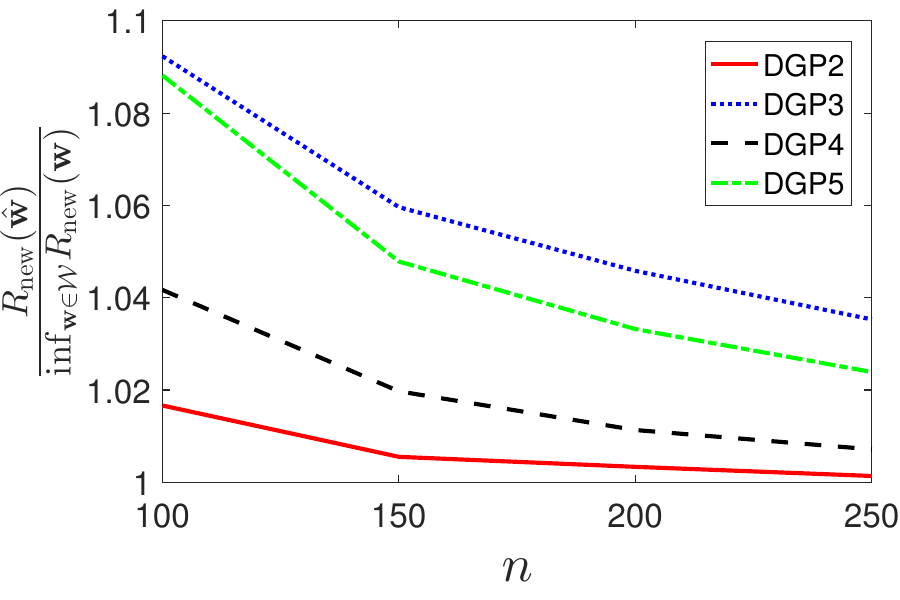}\label{fig:varifyTh3a}
	}
	\subfigure[The trend of $\hat R_1$ under DGP2]{
		\includegraphics[width=7cm]{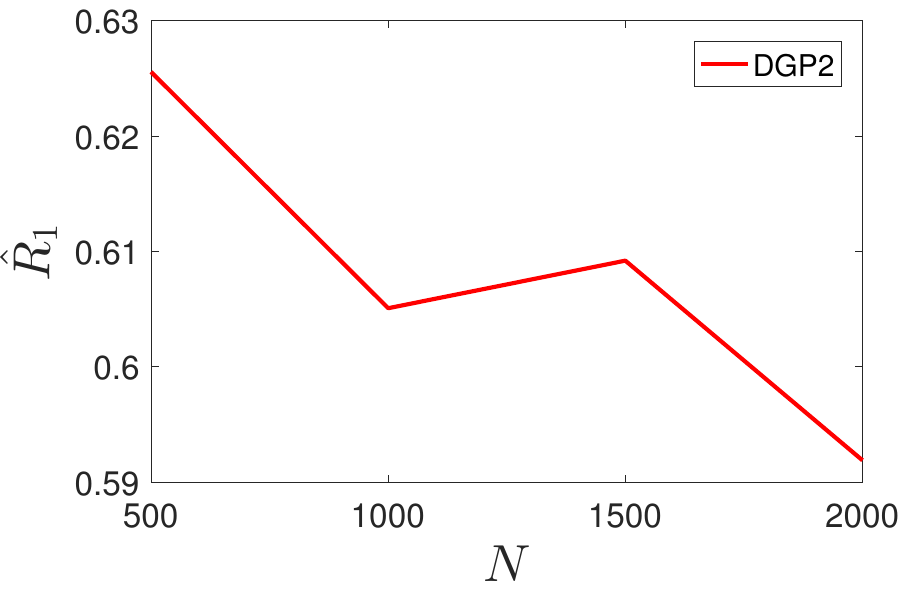}\label{fig:varifyTh3b}
	}\\
\subfigure[The trend of $\hat R_1$ under DGP3]{
	\includegraphics[width=7cm]{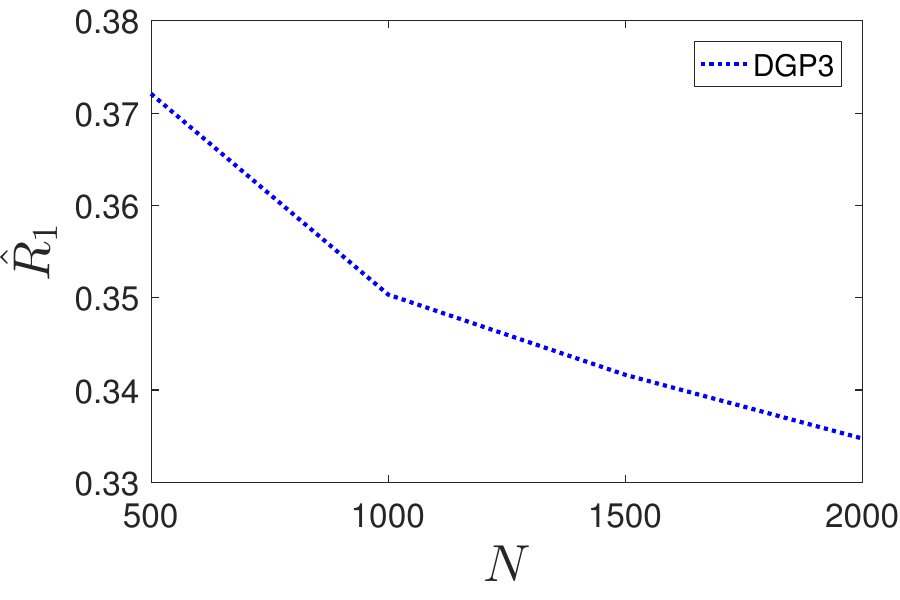}
}\label{fig:varifyTh3c}
\subfigure[The trend of $\hat R_1$ under DGP4 and DGP5]{
	\includegraphics[width=7cm]{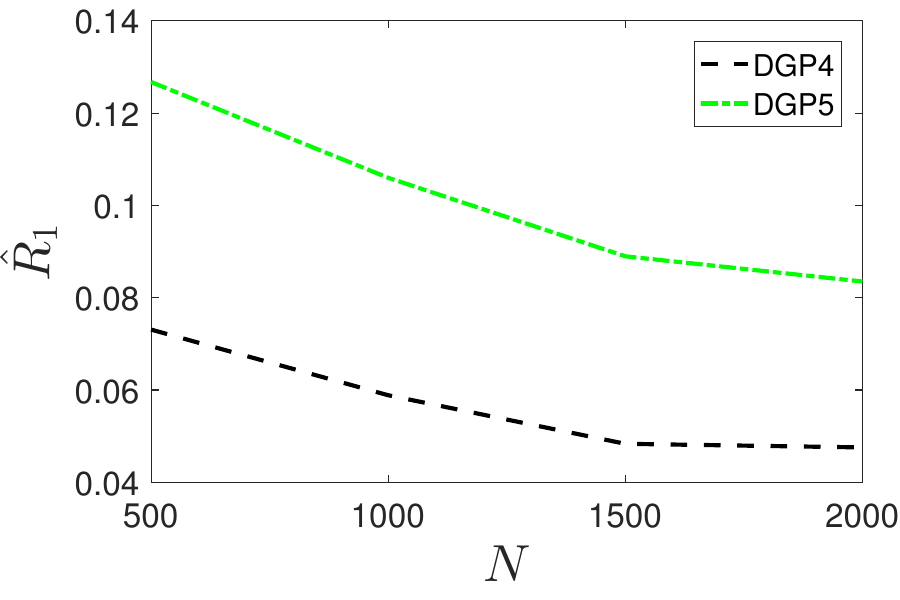}\label{fig:varifyTh3d}
}
	\caption{Verification of Theorem \ref{th:opt} under DGP2 to DGP5.} \label{fig:varifyTh3}
\end{figure}

We set $p=4$ and $N=20n$ to observe the trend of $R_\new(\hat\w)/\inf_{\w\in\calW}R_\new(\w)$ corresponding to the 5-CRMA method changes with $n$ under DGP2 to DGP5.
The results in Subfigure \ref{fig:varifyTh3a} shown are the mean values of 500 repetitions for each $n$. We define 
\be
\hat R_1=\left\{\left.\frac{R_\new(\hat\w)}{\inf_{\w\in\calW}R_\new(\w)}\right|_{N}-1\right\}\times\left\{\left.\frac{R_\new(\hat\w)}{\inf_{\w\in\calW}R_\new(\w)}\right|_{N=0}-1\right\}^{-1}\n
\ee
as the indicator for comparing the convergence rate of asymptotic optimality between the 5-CRMA and 5-LABEL methods. According to Theorem \ref{th:opt}, when $N< M^2n-n$, we have $\hat R_1=O_p\left(\sqrt{n/(n+N)}\right)$. We set $p=4$ and $n=100$ to observe the trend of $\hat R_1$ with changes in $N$. The results can be seen in Subfigures \ref{fig:varifyTh3b}-\ref{fig:varifyTh3d}.

From Subfigure \ref{fig:varifyTh3a}, it can be seen that in DGP2 to DGP5, as $n$ increases, the ${R_\new(\hat\w)}/{\inf_{\w\in\calW}R_\new(\w)}$ corresponding to our method tends to 1, which is consistent with the conclusion of Theorem \ref{th:opt}. 
In Subfigures \ref{fig:varifyTh3b}-\ref{fig:varifyTh3d}, $\hat R_1$ decreases as $N$ increases. This indicates that the asymptotic optimality convergence rate of the 5-CRMA method is faster than that of the 5-LABEL method, i.e., the inclusion of unlabeled data accelerates the convergence rate of asymptotic optimality. In summary, the results in Figure \ref{fig:varifyTh3} validate the conclusion of Theorem \ref{th:opt}.

\begin{figure}[htbp]
	\centering
	\subfigure[Weight consistency verification]{
		\includegraphics[width=7cm]{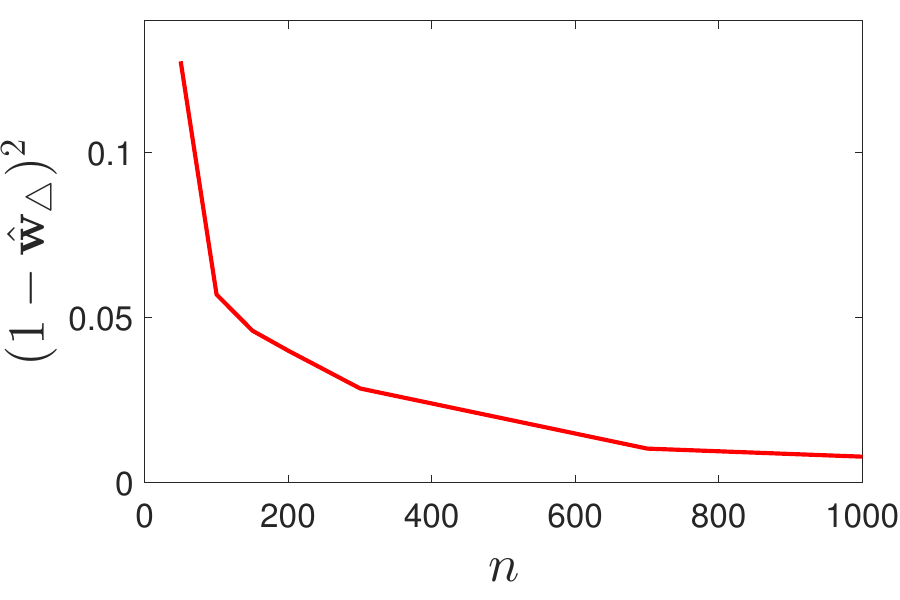}\label{fig:varifyTh4a}
	}
	\subfigure[The trend of $\hat R_2$ under DGP1]{
		\includegraphics[width=7cm]{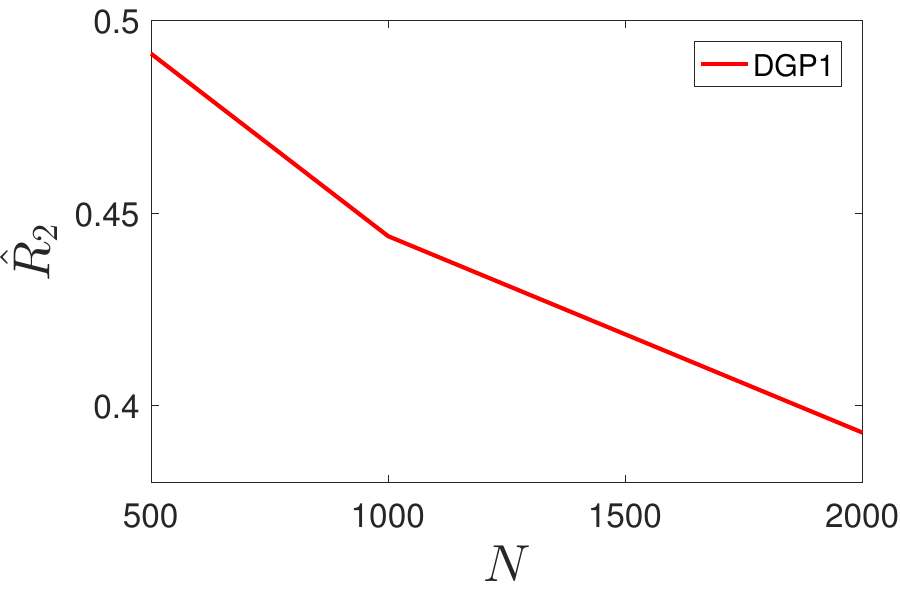}\label{fig:varifyTh4b}
	}
	\caption{Verification of Theorem \ref{th:weight} under DGP1.} \label{fig:varifyTh4}
\end{figure}

Similar to the verification of Theorem \ref{th:opt}, we set $p=4$ and $N=20n$ under DGP1 to observe how $(1-\hat\w_\triangle)^2$ corresponding to the 5-CRMA method changes with $n$. The results in Subfigure \ref{fig:varifyTh4a} shown are the mean values of 500 repetitions for each $n$. Define 
\be
\widehat R_2=\frac{\left.(1-\widehat \w_\triangle)^2\right|_{N}}{\left.(1-\widehat \w_\triangle)^2\right|_{N=0}}\n
\ee
as the indicator for comparing the convergence rate of weight consistency between the 5-CRMA and 5-LABEL methods. We set $p=4$ and $n=100$ to observe the trend of $\hat R_2$ with changes in $N$. The results can be seen in Subfigure \ref{fig:varifyTh4b}.

In Figure \ref{fig:varifyTh4}, $(1-\hat\w_\triangle)^2$ and $\hat R_2$ decrease with $n$ and $N$, respectively. This indicates that the sum of the weights corresponding to the correct candidate models converges to 1, and the inclusion of unlabeled data accelerates the convergence rate of weight consistency. Thus, Theorem \ref{th:weight} is validated.

\subsection{Simulated Los Angeles homeless dataset analysis}
In this subsection, we use a simulated data version of the Los Angeles homeless dataset provided by \cite{song2023general} to verify the performance of the proposed model averaging estimator on prediction.

Los Angeles County has 2054 census tracts, of which 244 ``hot tracts" with a large number of homeless people were counted. 
Further, 265 of the remaining 1810 tracts were randomly selected and the homeless people were counted, leaving 1545 tracts to be predicted. 
We use \emph{Perc.Industrial}, \emph{Perc.Residential}, \emph{Perc.Vacant}, \emph{Perc.Commercial}, \emph{Perc.OwnerOcc}, \emph{Perc.Minority}, \emph{MedianHouseholdIncome} ($p=7$) from the U.S. Census as covariates recommended by \cite{Zhang2019semi} and \cite{song2023general}, and the response $Y$ is the count of homeless people in each tract.
Removing 244 ``hot tracts", we have $n=265$ and $N=1545$ in this semi-supervised problem.
The count of homeless people in the $N$ tracts is missing in practice, and we use the same delete-one CV method as in \cite{Hohsuk2013copula} to evaluate the prediction performance of all methods on $n$ tracts:
\be
\mathrm{E}_1^{\text{mean}}=\frac{1}{n}\sumi\left|Y_i-\hat\mu_{-i}(\X_i)\right|,&\qquad& \mathrm{E}_1^{\text{median}}=\text{median}_{1\leq i\leq n}\left|Y_i-\hat\mu_{-i}(\X_i)\right|,\n\\
\mathrm{E}_2^{\text{mean}}=\frac{1}{n}\sumi\left\{Y_i-\hat\mu_{-i}(\X_i)\right\}^2,&\qquad& \mathrm{E}_2^{\text{median}}=\text{median}_{1\leq i\leq n}\left\{Y_i-\hat\mu_{-i}(\X_i)\right\}^2,\n
\ee
where $\hat\mu_{-i}(\X_i)$ is the estimator of $\mu(\X_i)$ obtained by each method using data $\{(Y_j,\X_j):j\neq i,j=1,\ldots,n\}\cup\{\X_j:j=n+1,\ldots,n+N\}$. The simulation results are given in Table \ref{tab:Homeless}. Since $K$ has a minor impact on our proposed method, only the results for $K = 5$ are reported later.
Table \ref{tab:Homeless} suggests that the 5-CRMA method produces the smallest error under all four delete-one CV versions.
In terms of predicting the count of homeless people in each tract, our proposed method yields the most accurate predictions among all methods.

\begin{table}[htbp]
	\setlength{\tabcolsep}{4mm}
	\centering
	\caption{Comparison of CV errors for different methods.}\label{tab:Homeless}
	\begin{tabular}{ccrrrrrr}
		\hline
		& \multicolumn{1}{c}{5-CRMA}  & \multicolumn{1}{c}{PSSE} & \multicolumn{1}{c}{PI} & \multicolumn{1}{c}{EASE} & \multicolumn{1}{c}{BICMS} & \multicolumn{1}{c}{SBIC} & \multicolumn{1}{c}{EWMA} \\
		\hline
		$\mathrm{E}_1^{\text{mean}}$ & \textbf{2.80}     & 3.15  & 3.09  & 3.07    & 3.32   & 3.32  & 3.14   \\
		$\mathrm{E}_1^{\text{median}}$ 
		&\textbf{2.52}       & 2.68  & 2.63  & 2.55           & 2.82   & 2.82  & 2.59  \\
		$\mathrm{E}_2^{\text{mean}}$ & \textbf{12.95}   & 16.05 & 16.02 & 16.05    & 18.28 & 18.40 & 16.48  \\
		$\mathrm{E}_2^{\text{median}}$ & \textbf{6.34}    & 7.17  & 6.92  & 6.49    & 7.95   & 7.96  & 6.70 \\\hline
	\end{tabular}%
	\begin{tablenotes}
		\footnotesize
		\item[1] Note: The smallest error in each row is in boldface type.
	\end{tablenotes}
\end{table}%

\section{California housing data analysis}\label{sec:semi_realdata}
In this section, the California housing dataset from the KEEL-dataset repository is used for experiments. The task of this dataset is to predict the median house value for each block based on information from the U.S. Census about all the block groups in California. 
In this dataset, $Y$ is the median house value and $\X$ contains \emph{Longitude}, \emph{Latitude}, \emph{HousingMedianAge}, \emph{TotalRooms}, \emph{TotalBedrooms}, \emph{Population}, \emph{Households}, and  \emph{MedianIncome}.

This dataset contains 20,640 observations. As it is costly to collect the median house value in each block group, we consider the semi-supervised setting. 
From the 20,640 observations $n=\{200,500\}$ observations are randomly selected without replacement as labeled data $\calL$, $N=\{n,2n,4n,6n,8n\}$ observations as unlabeled data $\calU$, and $n+N$ observations as test set $\mathcal{D}_{\text{test}}$. 
These settings give rise to 10 combinations of $(n,N)$, and the process of randomly selecting observations is repeated 500 times under each combination. 

\begin{figure}[htbp]
	\centering
	\subfigure[$n=200$]{
		\includegraphics[width=7cm]{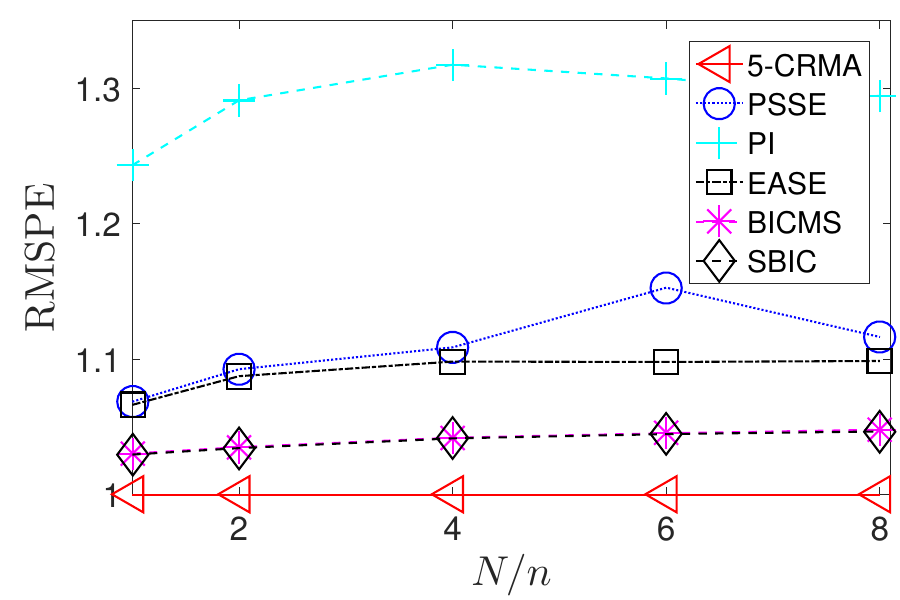}
	}
	\subfigure[$n=500$]{
		\includegraphics[width=7cm]{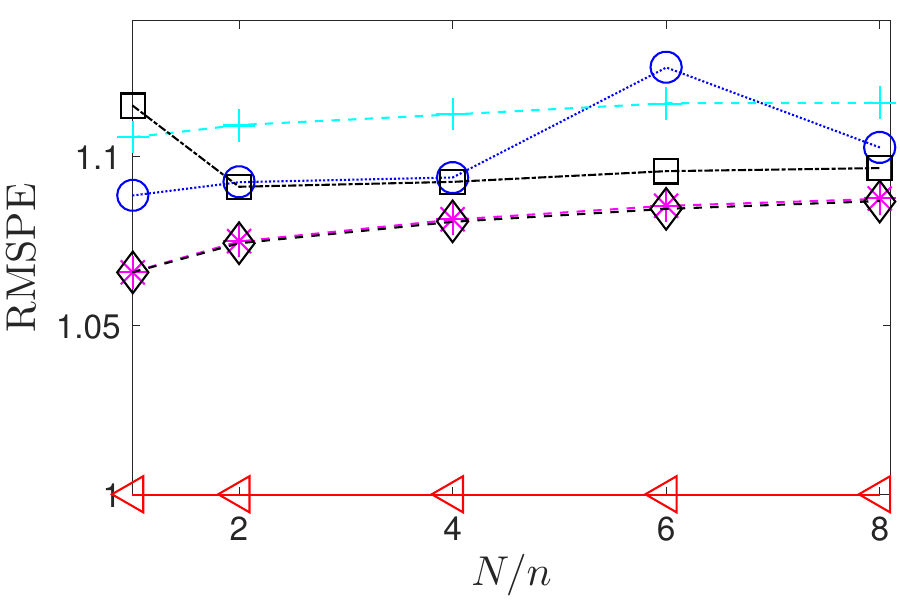}
	}
	\caption{Comparison of MSPEs for different methods.} \label{fig:California}
\end{figure}

The candidate copulas are the same as those in subsection \ref{sec:simulation}.
For comparison purposes, we divide the MSPE of all methods by the MSPE of the 5-CRMA method, so the MSPE of the 5-CRMA method is equal to 1. In addition, we delete the results of the EWMA method because the MSPE of the EWMA method is very large.
The mean values of 500 relative MSPEs (RMSPE) for each combination of $(n,N)$ are displayed in Figure \ref{fig:California}. Figure \ref{fig:California} indicates that except for the 5-CRMA method, the RMSPEs of all methods are greater than 1, that is, the 5-CRMA method generates the least error in all cases. 

\section{Concluding remarks}\label{sec:semi_remarks}
In this paper, we establish model averaging regression prediction within the semi-supervised learning framework and discuss the role of unlabeled data from the perspectives of the convergence rates of asymptotic optimality and weight consistency. Our model averaging estimator achieves faster convergence rates than the supervised counterpart.
Simulation studies and the California housing dataset indicate that our model averaging estimator exhibits a clear competitive advantage in prediction compared to current advanced semi-supervised learning methods.

There are two promising directions for future research: (1) Analysis of semi-supervised learning in high-dimensional settings, exploring the effects of exploiting unlabeled data in such scenarios; (2) Addressing the issue of covariate shift, where the covariate distribution of labeled data differs from that of unlabeled data, and investigating effective ways to leverage unlabeled data in such cases.


\section*{Supplementary material}
Supplementary material is available online. 

\baselineskip=16pt
\bibliographystyle{sty}
\bibliography{RefList}

\end{document}